# Structures and magnetic properties of Co-Zr-B magnets studied by first-principles calculations

Xin Zhao, Liqin Ke, Manh Cuong Nguyen, Cai-Zhuang Wang[*], and Kai-Ming Ho[§]

Ames Laboratory, US DOE and Department of Physics and Astronomy, Iowa State University, Ames, Iowa 50011, USA

**Abstract**

The structures and magnetic properties of the Co-Zr-B alloys near the $Co_5Zr$ composition were studied using adaptive genetic algorithm and first-principles calculations to guide further experimental effort on optimizing their magnetic performances. Through extensive structural searches, we constructed the contour maps of the energetics and magnetic moments of the Co-Zr-B magnet alloys as a function of composition. We found that the Co-Zr-B system exhibits the same structural motif as the "$Co_{11}Zr_2$" polymorphs, which plays a key role in achieving high coercivity. Boron atoms can either substitute selective cobalt atoms or occupy the interstitial sites. First-principles calculation shows that the magnetocrystalline anisotropy energies can be significantly improved through proper boron doping.

[*]e-mail: wangcz@ameslab.gov

[§]e-mail: kmh@ameslab.gov

**Introduction**

As promising candidates for non-rare-earth permanent magnets, $Co_xZr$ alloys with x near 5 and related compounds, such as Co-Zr-B, Co-Zr-M-B (M=C, Si, Mo, etc.), have attracted considerable attentions [1-10]. Great effort has been devoted to the improvement of their hard magnetic properties. The reported highest coercivity was 9.7 kOe, found in annealed $Co_{74}Zr_{16}Mo_4Si_3B_3$ ribbons [8] and the optimal magnetic properties were obtained in $Co_{80}Zr_{18}B_2$ with intrinsic coercivity Hc=4.1 kOe and energy product $(BH)_{max}$ = 5.1 MGOe [5]. More recently, cluster beam deposition has been used to make Co-Zr/Hf samples and energy products of 16-20 MGOe were reported [9, 10]. The Co-Zr/Hf magnet alloys typically contain multiple phases and identifying the phase responsible for the magnetic hardness has been one of the research focuses. Several studies [11-13] assumed that the hard magnetic phase in the Co-Zr system is the metastable $Co_5Zr$ phase with the structure of $Ni_5Zr$. However, $Ni_5Zr$ structure is cubic and thus unlikely to provide strong magnetocrystalline anisotropy energies, which was confirmed by first-principles calculations [14]. $Co_3ZrB_2$ has also been proposed to be a candidate for the hard magnetic phase [15], which remains to be validated.

Determining the hard magnetic phase in the above-mentioned alloys has been a long-standing issue due to the ambiguity of their crystal structures. Recently, progress has been made in solving the crystal structures of the complex $Co_xZr$ alloys. Using adaptive genetic algorithm (AGA) [16], we studied the crystal structures of the rhombohedral, hexagonal, and orthorhombic polymorphs close to the $Co_{11}Zr_2$ intermetallic compound

[14]. The common building block in the structures of these polymorphs was identified as a derivative from the SmCo$_5$ structure. Decrease of the temperature induces a phase transition from high symmetry rhombohedral/hexagonal phase to low symmetry orthorhombic/monoclinic phase, along with a slight increase of the Co concentration. The experimental data from the x-ray diffraction (XRD) and transmission electron microscopy were well explained by the crystal structures obtained from AGA searches. Through first-principles magnetic properties calculations, the hard magnetic phase in the Co$_x$Zr alloys was identified to be the high-temperature rhombohedral/hexagonal phase.

In this work, we extended the investigation to the effect of boron doping on the structures and magnetic properties of the Co$_x$Zr alloys. Structure searches by AGA allowed us to access the preferred positions of boron atoms, thus energetics and magnetic properties of different Co-Zr-B compositions can be studied by first-principles calculations.

**Computational methods**

Crystal structures of Co-Zr-B were investigated by AGA [14, 16]. The structure searches were performed without any assumption on the Bravais lattice type, atom basis or unit cell dimensions. The size of the unit cell studied in this work was up to 100 atoms. In the AGA search for this system, embedded-atom method [17] was used as the auxiliary classical potential. The parameters for Co-Co and Zr-Zr interactions were from the literature [18]. B-B interaction and the crossing-pair interactions (*i.e.*, B-Co, B-Zr and Co-Zr) were modeled by Morse function (eq. 1), with 3 adjustable parameters each (D, α,

$r_0$). For Co and Zr atoms, parameters of the density function and embedding function were taken from ref. [18] as well, and for B atom, exponential decaying function was used as the density function (eq. 2, with 2 adjustable parameters: α, β) and the form proposed by Benerjea and Smith [19] was used as the embedding function (eq. 3, with 2 adjustable parameters: $F_0$, γ).

$$\phi(r_{ij}) = D\left[e^{-2\alpha(r_{ij}-r_0)} - 2e^{-\alpha(r_{ij}-r_0)}\right] \quad \text{(eq. 1)}$$

$$\rho(r_{ij}) = \alpha \exp[-\beta(r_{ij} - r_0)] \quad \text{(eq. 2)}$$

$$F(n) = F_0[1 - \gamma \ln n]n^\gamma \quad \text{(eq. 3)}$$

The total energy of the system then has the following form:

$$E_{total} = \frac{1}{2}\sum_{i,j(i\neq j)}^{N} \phi_{ij}(r_{ij}) + \sum_i F_i(n_i) \quad \text{(eq. 4)}$$

Where $r_{ij}$ is the distance between atoms i and j, $n_i = \sum_{j\neq i} \rho_j(r_{ij})$ is the electron density at the site occupied by atom i.

The potential parameters were adjusted adaptively by fitting to the DFT energies, forces, and stresses of selected structures according to the AGA scheme [16]. The fitting was performed by the force-matching method with stochastic simulated annealing algorithm implemented in the *potfit* code [20, 21]. First-principles calculations were performed using the projector augmented wave (PAW) method [22] within density functional theory (DFT) as implemented in VASP code [23]. The exchange and correlation energy is treated within the spin-polarized generalized gradient approximation (GGA) and parameterized by Perdew-Burke-Ernzerhof formula (PBE) [24]. Wave functions are expanded in plane waves up to a kinetic energy cut-off of 350 eV. Brillouin-zone integration was performed using the Monkhorst-Pack sampling scheme [25] over $k$-point mesh resolution of $2\pi \times 0.033$ Å$^{-1}$. The formation energy $E_F$ of the alloy is calculated as:

$$E_F = [E(Co_m Zr_n B_p) - m \cdot E(Co) - n \cdot E(Zr) - p \cdot E(B)]/(m+n+p) \quad (eq.\ 5)$$

Where $E(Co_m Zr_n B_p)$ is the total energy of the $Co_m Zr_n B_p$ alloy; $E(Co)$, $E(Zr)$ and $E(B)$ are the energy per atom of Co, Zr and B in the reference structures, which are HCP Co, HCP Zr and α-boron, respectively.

Intrinsic magnetic properties of the Co-Zr-B structures, such as magnetic moment and magnetocrystalline anisotropy energy (MAE) were calculated using VASP code. The spin-orbit coupling (SOC) is included using the second-variation procedure [26]. We also calculated the MAE of the rhombohedral Co$_5$Zr structure by carrying out all-electron calculations using the full-potential (FP) LMTO method to check VASP calculation results. In addition, by evaluating the SOC matrix elements, the anisotropy of orbital moment and MAE was resolved into sites, spins and orbital pairs [27] to identify their contribution to the magnetic properties. Curie temperature ($T_c$) is checked for selected structures using mean-field approximation and more details can be found in Ref. [28].

**Results and discussions**

To validate the selection of the auxiliary classical potential, we first performed crystal structure search for the $Co_3ZrB_2$ phase, whose crystal structure was well-characterized. The ground state structure of $Co_3ZrB_2$ was successfully found in the AGA search with above setup [16]. Further calculations on its magnetic properties by DFT showed this phase is non-magnetic with zero magnetic moments. Therefore, this structure cannot be responsible for the magnetic properties observed in the Co-Zr-B system.

In order to obtain practically useful magnets, we then performed extensive AGA searches for Co-Zr-B with Co:Zr ratio around 5 and boron composition less than 6 at %. The contour map of their formation energies is plotted in Fig. 1, where the compositions searched in current work are represented by squares. It can be seen that near the $Co_5Zr$ composition (Co, at % ~ 83.3%) there is a local minimum in the energy landscape, which explains the $Co_xZr$ (x ~ 5) phases obtained in experiments. For certain compositions at the high energy areas, such as $Co_{84}Zr_{15}B$, and $Co_{46}Zr_8B_2$, it is unlikely to synthesize such compounds experimentally. Among the compositions considered in Fig. 1, the lowest formation energy is found around $Co_{40}Zr_9B$ and $Co_{40}Zr_8B_2$, which are consistent with experimental results since most of samples produced by experiments were around these compositions [4-6]. In the following, structures and magnetic properties of the Co-Zr-B alloys will be discussed respectively.

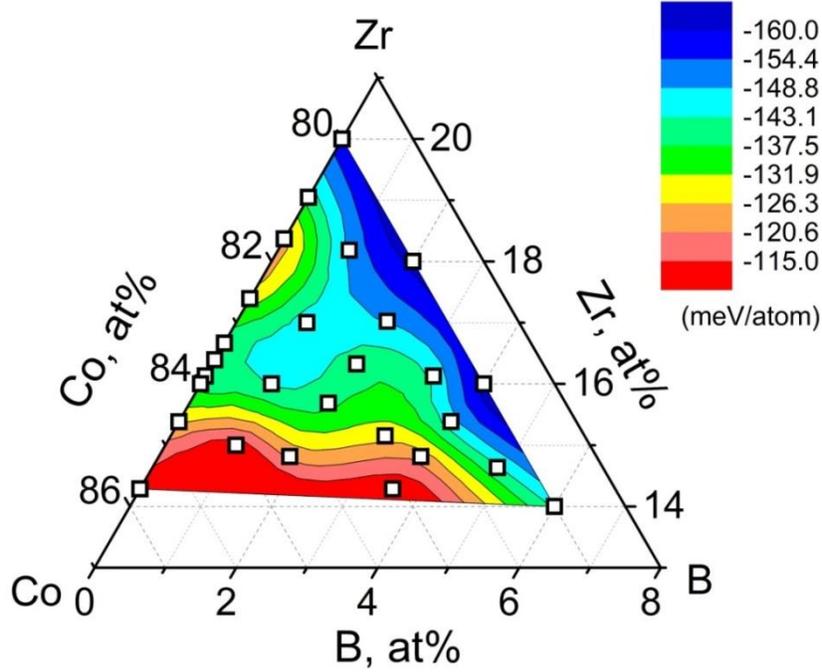

FIG. 1. Contour map of the formation energies in the Co-Zr-B system. Only partial composition range is considered and the squares represent the compositions searched by AGA in the present work.

**I. Structures**

Several low-energy boron-doped $Co_xZr$ structures obtained from our AGA searches are plotted in Fig. 2(a-d). Co and Zr atoms form the same building block as discovered in $Co_xZr$ [14], while boron atoms can either substitute Co atoms [e.g. Fig. 2(a)] or occupy interstitial positions [e.g. Fig. 2(c)] in company with local distortions. In the $Co_{40}Zr_8B_2$ structure plotted in Fig. 2(b), boron atoms can be considered as interstitial atoms in the $Co_5Zr$ structure of high temperature phase, or as substitutional atoms in the $Co_{5.25}Zr$ structure of low temperature phase, because the main difference between the $Co_5Zr$ and

Co$_{5.25}$Zr structures comes from the different packing density of one of the two pure Co layers [14]. In our previous study, we also showed the layer-stacking feature in Co$_x$Zr polymorphs is frequently interrupted to adjust the strain due to the rippled hexagonal Co layer [14]. Figure 2(d) shows a similar structure with boron atoms located at the interruption site.

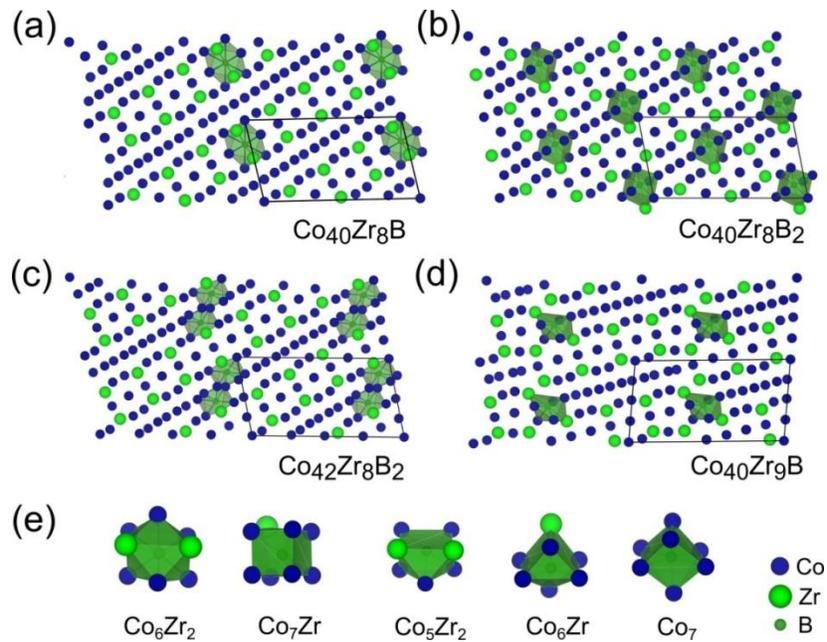

FIG. 2. Examples of the low-energy structures obtained from the AGA searches with compositions of (a) Co$_{40}$Zr$_8$B; (b) Co$_{40}$Zr$_8$B$_2$; (c) Co$_{42}$Zr$_8$B$_2$; (d) Co$_{40}$Zr$_8$B. Unit cell of each structure is indicated by black lines. (e) Typical boron-centered clusters extracted from the Co-Zr-B crystal structures. The label under each cluster represents the neighboring atoms of boron.

To give a better picture of the local environments of boron atoms, Fig. 2(e) listed several typical boron-centered clusters found in the Co-Zr-B structures. In general, the nearest neighbor distances for the B-Co and B-Zr pairs are about 2.1 Å and 2.6 Å, respectively. The coordination number of the boron atoms is 7 or 8, and the neighboring atoms are found to be Co or Zr atoms in most cases. The effect of different boron positions on the magnetic properties will be discussed later.

The structure and glass formability in the Co-Zr-B alloy system have been studied experimentally [29-31]. In the XRD analysis [29], the intensity of the crystalline peaks becomes weaker and broader as the boron content increases, indicating the reduction of the crystalline size in the samples. Amorphous and partially crystalline alloys have also been observed in this system [2, 31]. From the AGA search, we found that all the low-energy structures of Co-Zr-B have low symmetries (triclinic system) due to the distortions induced by the doping of boron atoms. Moreover, many different structures were found to have closely competitive energies (within a few meV per atom), similar to the $Co_xZr$ binary system. Therefore, within the composition range plotted in Fig. 1, growing single crystals of Co-Zr-B alloy is difficult and the sample is expected to have small grains and defects.

## II. Magnetic properties

### A. High-temperature $Co_5Zr$ phase revisited

In our previous study [14], the high-temperature rhombohedral phase was assigned to be responsible for the magnetic hardness in the $Co_xZr$ alloys. The full potential calculation

(GGA) showed it has a magnetic moment of around 1.0 $\mu_B$/atom and MAE of 1.4 MJ/m$^3$. The rhombohedral structure, plotted in Fig. 3(a), has a space group R32 (#155) and 4 inequivalent Co sites as indicated by different colors in Fig. 3. Views along c axis of the different layers are plotted in Fig. 3(b). Among the four inequivalent Co sites, two of them (Co1, Co3) have nine-fold multiplicity and the other two (Co2, Co4) have six-fold multiplicity.

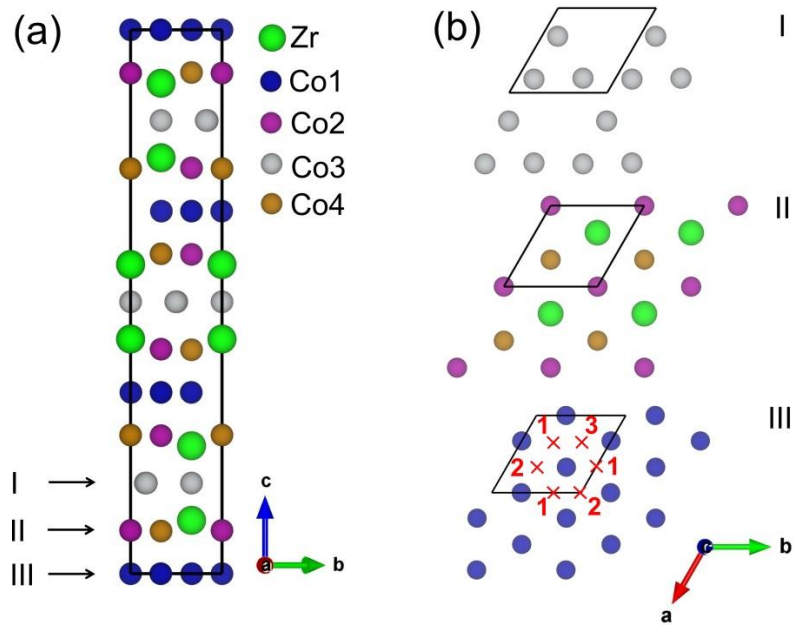

FIG. 3. (a) Crystal structures of the rhombohedral Co$_5$Zr with different Co sites presented by different colors. The lattice parameters of the structure are a=4.66 Å and c=24.0 Å. It has one Zr site: 6c (0.0000, 0.0000, 0.4314) and four Co sites: Co1 9d (0.3300, 0.0000, 0.0000), Co2 6c (0.0000, 0.0000, 0.0795), Co3 9e (0.4946, 0.0000, 0.5000), and Co4 6c (0.0000, 0.0000, 0.2549). (b) Views of layer I, II and III along **c** axis. In the plot of layer III, all possible interstitial positions are grouped into 3 inequivalent sites based on symmetry. Unit cells of the crystal structures are indicated by the black boxes.

To examine the contribution of different sites to the magnetic properties of the rhombohedral phase, Fig. 4 shows the variations of orbital magnetic moments and relativistic energy as functions of the spin rotation. Rhombohedral $Co_5Zr$ has uniaxial anisotropy. By evaluating the SOC matrix element, we found the Co3 site has in-plane magnetic easy axes while all other Co sites, especially Co4, support the uniaxial anisotropy. As shown in Fig. 4, the correlation between orbital moment and magnetic anisotropy is obvious. Co1, Co2 and Co4 sites have larger orbital magnetic moments along the z axis while Co3 has larger orbital magnetic moments when spin is along in-plane directions. The MAE calculated in LDA is smaller than the one calculated using PBE functional [14]. By evaluating the SOC matrix elements, we found that this difference mostly comes from the Co1 site, whose contribution to MAE nearly disappears in LDA. The MAE contributions from all other sites barely depend on the exchange-correlation functionals used in our calculations. Above analysis indicates if the Co3 site can be modified, such as substituting Co3 by other elements, the MAE of the system could be improved.

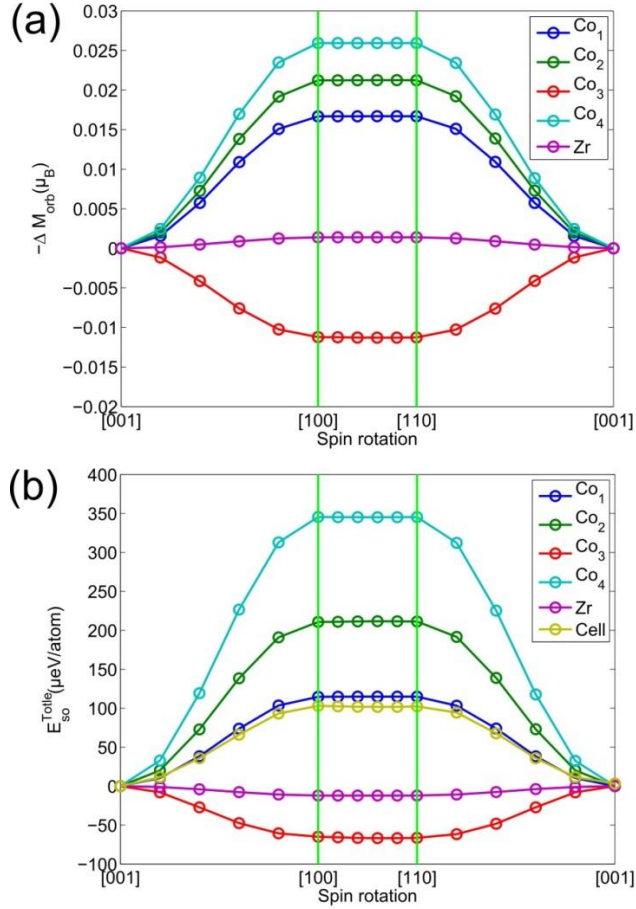

FIG. 4. Variations of the orbital moment (a) and relativistic energy (b) as a function of spin quantization direction. Orbital moment and relativistic energy values are averaged over all atoms which belong to the corresponding inequivalent sites.

**B. Boron doping on magnetic moments**

To map out the magnetic moments of the Co-Zr-B alloys, results from VASP calculations were collected for all the compositions presented in Fig. 1. The results of total moments in the system are plotted in Fig. 5(a), and the partial contributions from Co, Zr, and B atoms are plotted in Fig. 5(b), (c), and (d) respectively. The total magnetic moment per atom is calculated as the moment of the whole system divided by the total number of atoms, while the moment contribution from atom type M (M=Co, Zr, or B) is calculated as the moment from all the M atoms divided by the number of atom M. Results plotted in Fig. 5 for each composition were averaged over ten lowest-energy structures from the AGA searches.

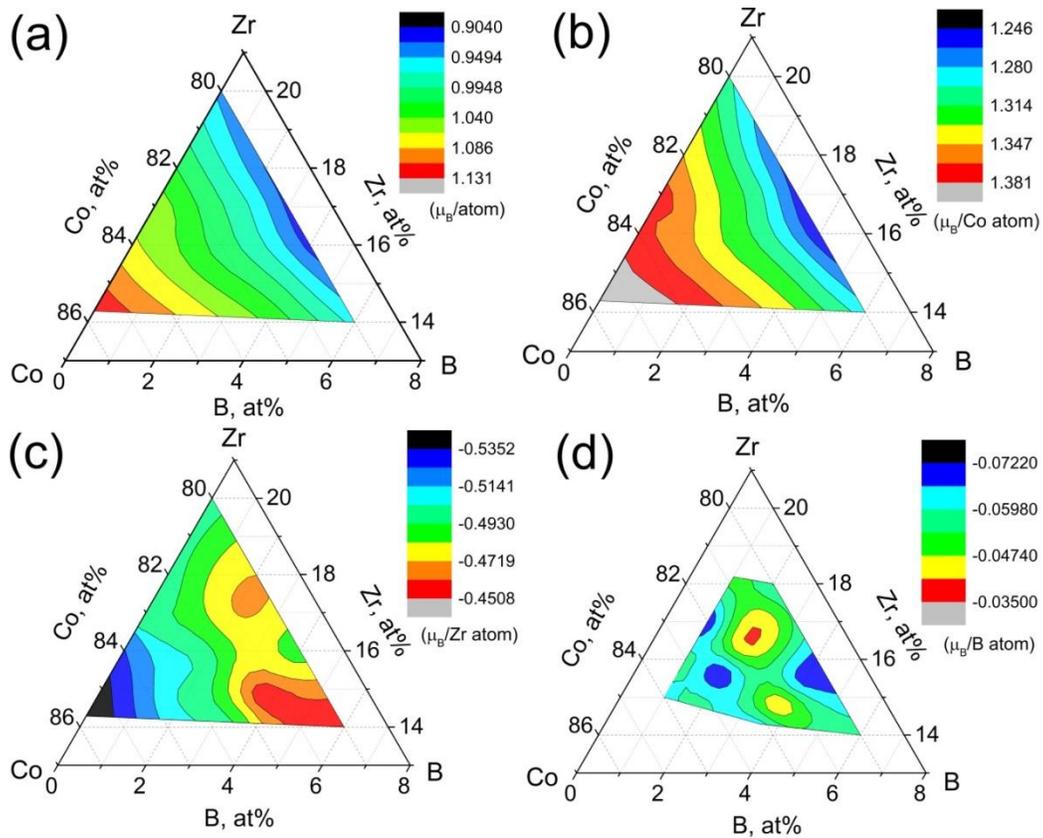

FIG. 5 (a) Contour map of the total magnetic moment per atom in the composition range studied for Co-Zr-B; (b, c, d) contour plots of the partial contributions from Co, Zr and B atoms to the magnetic moment respectively.

It can be seen that the magnetization in the Co-Zr-B system mainly comes from the Co atoms. Both the Zr and B atoms are antiferromagnetically coupled to the Co atoms. As shown in Fig. 5(a), the magnetic moment of the system becomes smaller with the decrease of Co atomic composition, which can be explained by two reasons. First, Zr and B atoms give negative contribution to the total moment of the system. More Zr and B atoms will lower the moment of the system. Second, the Zr and B atoms suppress the moment of Co, which can be seen from Fig. 5(b). The average moment of the Co atoms is decreased with the increase of the Zr, B compositions. In contrast to the Co moment, the variation of the moment in Zr and B atoms as the function of composition is more complicated and there is no clear trend of how the magnetic moments of Zr and B change with composition. However, due to the small atomic percentages of the Zr and B atoms and their small moments, total magnetic moment of the system is dominated by Co atoms and varies in the same manner as that of Co.

**C. Boron doping on MAE**

The computational cost of calculating magnetocrystalline anisotropy energy can be enormous, which makes it infeasible to scan all the low-energy structures from AGA searches, especially when the unit cells contain as many as 100 atoms. In the following, the effect of boron doping on MAE was investigated based on the rhombohedral $Co_5Zr$ structure and the knowledge of the preferred sites by boron atoms from above analysis. All calculations were performed using VASP. To compare, the MAE of the rhombohedral structure calculated from VASP is about 1.6 MJ/m$^3$, which is very close to the result from FP calculations (1.4 MJ/m$^3$).

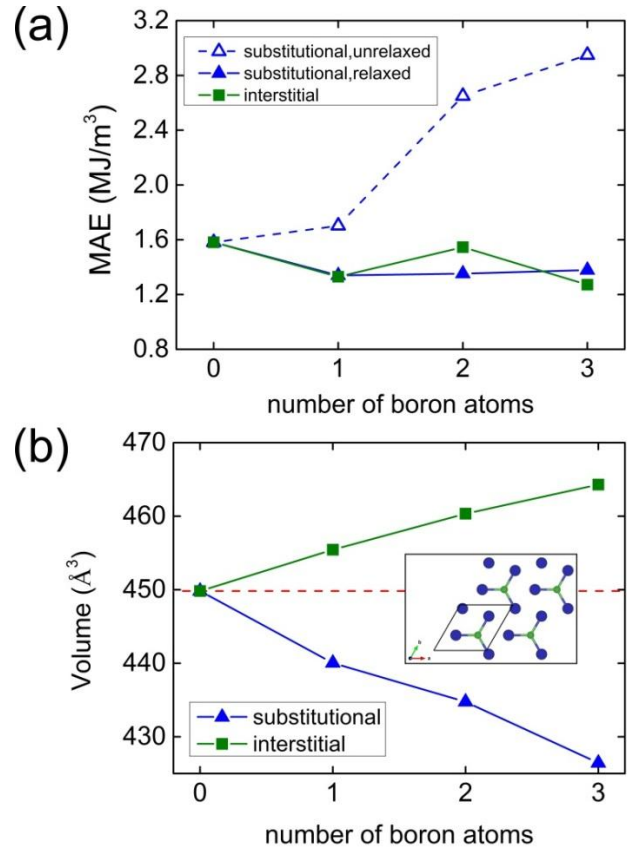

FIG. 6 (a) Effect of boron doping on MAE. The calculations were based on the rhombohedral Co$_5$Zr structure. In the case of substitution, results of both the relaxed (solid triangles) and unrelaxed structures (empty triangles) are plotted, while for interstitial positions, only results of the relaxed structures (solid squares) are plotted. The positions of doped boron atoms are discussed in the main text. (b) Volume comparison between the original Co$_5$Zr structure (dash line) and boron doped structures after DFT relaxations. The layer III with interstitial boron atom after relaxation is plotted as the inset.

We have showed that the same structure motif found in $Co_5Zr$ polymorphs also exists in the boron-doped $Co_xZr$ alloys, which explains the origin of the high coercivity observed in Co-Zr-B alloys. Referring to the rhombohedral $Co_5Zr$ structure plotted in Fig. 3, the boron atoms appear to prefer substituting Co atoms in layer I or entering layer III as interstitial atoms. Therefore, we scanned all the possibilities of adding up to 3 boron atoms into the conventional unit cell of the rhombohedral $Co_5Zr$ structure (Fig. 3a, 36 atoms) and selected the one with lowest energy for each scenario to calculate MAE. All structures have uniaxial anisotropy unless noted otherwise.

In the substitutional case, all the Co atoms in layer I belong to the same Wyckoff position, therefore the choices of substituting Co by B are limited. We found while substituting more than one Co atoms, lower energies were obtained with one B atom per layer, i.e. two B atoms substituting the same layer is not energetically favored for the 1×1 cell studied in this work. This can be explained by the fact that the size of the boron atom is much smaller than that of Co atom. Large distortions would be introduced if the boron density in one layer is too big. The calculated MAE results are plotted in Fig. 6(a) for both the relaxed and unrelaxed structures. It can be seen that the MAE increases significantly with the number of boron atoms for the unrelaxed structures, which confirms our speculation that replacing Co atoms at the Co3 site with other elements without modifying the structure can improve the MAE. However, after structure relaxations, MAE values of the boron substituted structures become slightly smaller than the original $Co_5Zr$ structure. From volume comparisons plotted in Fig. 6(b), we can see the relaxation changes the structure noticeably. The changes in the environments of Co atoms cause the change of their electronic configuration and contributions to the MAE.

In the interstitial case, there are three inequivalent positions in each Co1 layer where boron atoms can occupy, as indicated in Fig. 3(b). We scanned all the possibilities of adding up to 3 boron atoms and plotted the MAE results in Fig. 6(a). Considering that interstitial defects usually introduce much larger distortions to the neighboring atoms, we only calculated the MAE of the relaxed structures. We note again that adding one boron atom into each layer III gives more competitive energy and the relaxed structure of the B-Co mixed layer is plotted as the inset of Fig. 6(b). The MAE data shows the interstitial boron atoms in the Co1 layer do not change the MAE much compared with the high-temperature $Co_5Zr$ phase.

Although above calculations were performed on models that were created based on the rhombohedral $Co_5Zr$ structure, the results are representative due to the consideration of the preferable positions of boron. In our previous study, we showed in the low temperature $Co_{5.25}Zr$ phase where extra Co atoms packed in layer III to form the orthorhombic phase, the MAE is much lower than the high temperature $Co_5Zr$ rhombohedral phase [14]. However, if the extra atoms are boron atoms instead, such as Fig. 2(b), the MAE is expected to be close to the rhombohedral $Co_5Zr$ from above analysis. Meanwhile, when the density of boron substitution to the Co3 site is much smaller, such as Fig. 2(a), the distortion introduced to the neighboring atoms will be smaller, thus there exists a great chance to increase the anisotropy. Finally, we calculated the Curie temperatures for the model structures discussed above and it shows that the change in Curie temperature due to boron addition is not significant. The calculated Curie temperature is around 700 K which is high enough for practical use.

**Conclusion**

In summary, we studied the Co-Zr-B system using AGA method and first-principles calculations. We noted that the $Co_3ZrB_2$ phase is paramagnetic and cannot be responsible for magnetic hardness. Near the $Co_5Zr$ composition, the Co and Zr atoms in the structures of Co-Zr-B share the same structural motif as discovered in the $Co_xZr$ polymorphs, while boron atoms can appear both as substitutions for Co atoms or in the interstitial positions. Based on the AGA results, we constructed the formation energy and magnetic moment contour maps for partial composition range of the Co-Zr-B system, which can be used as guidance to adjust the experimental processing to further optimize the magnetic properties.

We believe the high coercivity observed in the ternary alloy system origins from the Co-Zr layer packing feature, as in the high temperature $Co_5Zr$ rhombohedral phase. Through the MAE calculations on Co-Zr-B model structures, we found both substitutional and interstitial boron atoms give similar magnetic anisotropy energies as the original rhombohedral $Co_5Zr$ structure. Our calculations provide insight into the significant improvement of the MAE in Co-Zr system through chemical doping.

**Acknowledgement**

This work was supported by the US Department of Energy-Energy Efficiency and Renewable Energy, Vehicles Technology Office, PEEM program, and by the US Department of Energy, Basic Energy Sciences, Division of Materials Science and Engineering. The research was performed at the Ames Laboratory, which is operated for the U.S. DOE by Iowa State University under Contract No. DE-AC02-07CH11358. This research used resources of the Oak Ridge Leadership Computing Facility (OLCF) in Oak Ridge, TN and the National Energy Research Scientific Computing Center (NERSC) in Berkeley, CA.